\documentclass[reprint,amsmath,amssymb,aps,prl]{revtex4-2}

\usepackage{comment}

\usepackage{graphicx}
\usepackage{dcolumn}
\usepackage{bm}
\usepackage{braket}
\usepackage{amsmath}
\usepackage{mathtools}
\usepackage{subfig}
\usepackage{caption}
\begin{document}

\title{Missing Rung Problem in Vibrational Ladder Climbing}

\author{Takahiro Horiba}
\affiliation{Toyota Central Research and Development Labs., Inc., 41-1, Yokomichi, Nagakute, Aichi 480-1192, Japan}
\author{Soichi Shirai}
\affiliation{Toyota Central Research and Development Labs., Inc., 41-1, Yokomichi, Nagakute, Aichi 480-1192, Japan}\author{Hirotoshi Hirai}
\email{hirotoshih@mosk.tytlabs.co.jp}
\affiliation{Toyota Central Research and Development Labs., Inc., 41-1, Yokomichi, Nagakute, Aichi 480-1192, Japan}
\date{\today}

\begin{abstract}
We observed vanishing of the transition dipole moment, interrupting vibrational ladder climbing (VLC) in molecular systems. 
We clarified the mechanism of this phenomenon and present a method to use an additional chirped pulse to preserve the VLC.
To show the effectiveness of our method, we conducted wavepacket dynamics simulations for LiH dissociations with chirped pulses.
The results indicate that the efficiency of LiH dissociation is significantly improved by our method compared to conventional methods.
We also revealed the quantum interference effect behind the excitation process of VLC.
\end{abstract}

\maketitle

Controlling molecular reactions as desired is one of the ultimate goals of chemistry.
Compared to conventional macroscopic reaction control methods using temperature and pressure, 
proposed methods that directly control the quantum state of molecules 
using the electric field of a laser 
are expected to achieve dramatically improved efficiency and selectivity of reactions
\cite{brumer1986control,tannor1985control,baumert1991femtosecond}.
Such techniques, 
first proposed by Brumer and Shapiro in the 1980s \cite{brumer1986control}, 
are called ``coherent control'', 
and with the advent of femtosecond pulsed lasers and the development of laser shaping technology 
have been the subject of numerous studies 
and continue to attract considerable attention \cite{potter1992femtosecond,hikosaka2019coherent,goto2011strong}.
A promising coherent control method 
is vibrational ladder climbing (VLC) \cite{chelkowski1995adiabatic},
which employs a cascade excitation process in molecular vibrational levels
under chirped infrared laser pulses 
with time-dependent frequencies corresponding to the vibrational excitation energy levels.
VLC can focus an input laser pulse energy on a specific molecular bond.
Thereby, it is expected to realize highly efficient bond-selective photodissociation \cite{morichika2019molecular}.
VLC has been studied intensively both experimentally and theoretically.
Experimentally,
the vibrational excitation by VLC in diatomic molecules (NO \cite{maas1997vibrational}, HF \cite{marcus2006molecular}) 
and amino acids \cite{jewariya2010ladder} have been reported.
Recently, Morichika et al. succeeded in breaking molecular bonds 
in transition metal carbonyl WCO$_6$
with a chirped infrared pulse 
enhanced by surface plasmon resonance \cite{morichika2019molecular}.
Unfortunately, there are few experimental reports 
of successful bond breaking by VLC,
which remains a challenge.
In early theoretical works, Liu et al. and Yuan et al.
studied the classical motion of driven Morse oscillators \cite{liu1995nonlinear,liu1999classical,duan2000classical}.
They analyzed the excitation and dissociation dynamics 
of diatomic molecules under a chirped electric field 
using the action-angle variable, 
and found the condition for efficient excitation \cite{liu1995nonlinear,liu1999classical}.
Regarding the quantum aspects, 
a pioneering work by Friedland et al. \cite{marcus2004quantum}
focused on quantum and classical excitation processes, 
specifically quantum ladder climbing and classical autoresonance, 
and proposed parameters  
to characterize the quantum and classical phenomena \cite{marcus2004quantum,barth2011quantum}. 
Based on the characterization parameters, 
quantum and classical comparisons were made in various systems \cite{barth2014quantum,armon2017quantum,armon2019quantum,armon2020quantum} 
such as plasma systems \cite{barth2015ladder} and superconducting circuits \cite{shalibo2012quantum}.
The excitation condition proposed by Duan et al. 
and the characterization parameters proposed by Friedland et al. 
depend on the physical properties of the system 
such as the potential energy surface (PES) and the dipole function.
Thereby, these properties determine the characteristic of the VLC.
However, most of the theoretical treatments were 
based on simple model functions of these properties 
such as Morse potentials and linear dipole moments \cite{lin1998quantum,mishima1998theoretical,liu1995nonlinear,marcus2004quantum,witte2003controlling}.

\begin{figure*}[bt]
  \centering
  \subfloat[]{\includegraphics[keepaspectratio, scale=0.48]{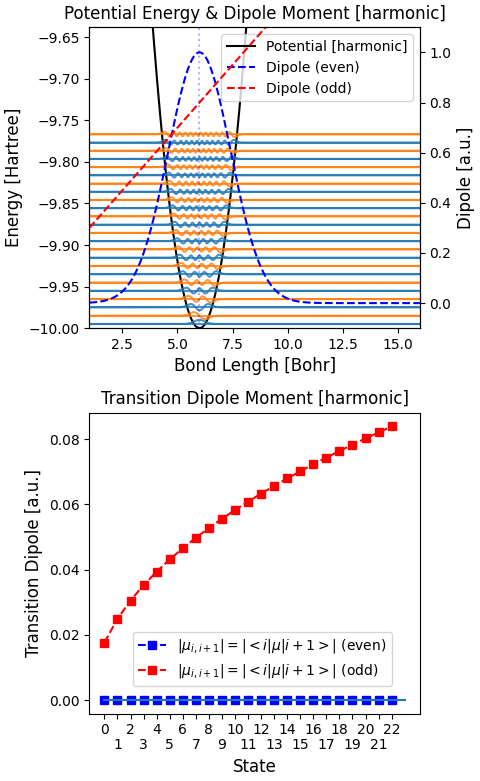}\label{fig:harm}}
  \subfloat[]{\includegraphics[keepaspectratio, scale=0.48]{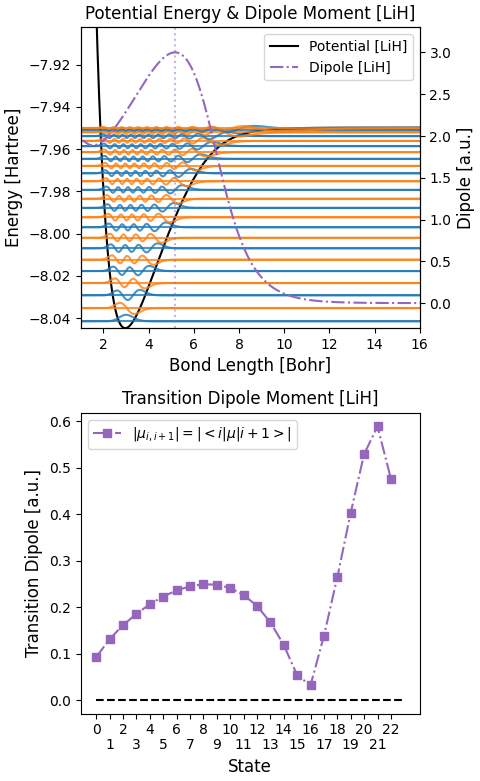}\label{fig:LiH}}
  \subfloat[]{\includegraphics[keepaspectratio, scale=0.48]{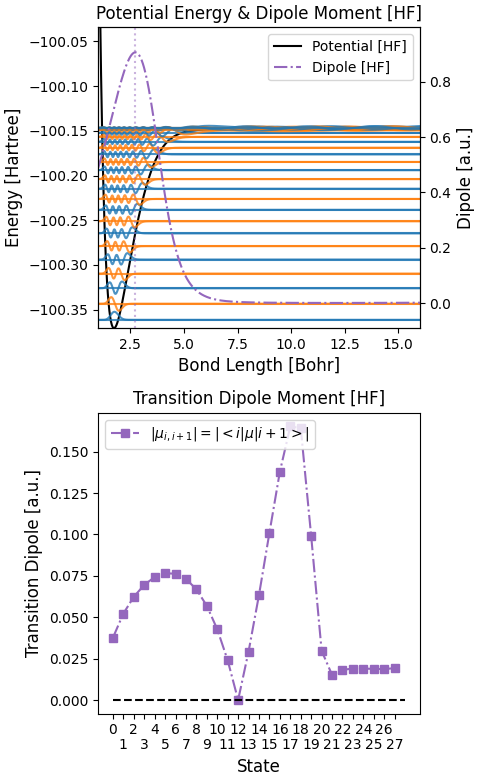}\label{fig:HF}}
  \caption{Upper panels: PESs and dipole moments for (a) the harmonic model, (b) LiH molecule, and (c) HF molecule. For the harmonic model, the model dipole moments are shown. 
  The orange and light blue lines shows the vibrational wavefunctions. 
  Lower panels:  absolute values of the corresponding TDMs.}
  \label{fig:pes_tdm}
\end{figure*}

In this study, 
we performed wavepacket dynamics simulations of VLC 
based on the PES and the dipole moment 
computed by a highly accurate \textit{ab-initio} quantum chemistry method.
The results of the simulations revealed a problem 
that has not previously been recognized in VLC: 
VLC is interrupted by the existence of a level 
with a nearly zero adjacent transition dipole moment (TDM), i.e. vanishing of the TDM.
We call this problem the missing rung problem (MRP) 
as it is like a rung missing in the middle of a ladder.
To clarify the issue in more detail, 
the absolute values of TDMs for the diatomic molecules LiH and HF are shown 
in the lower panels of Fig. \ref{fig:pes_tdm}\subref{fig:LiH} and \subref{fig:HF}, respectively.
It can be seen that 
the TDMs of the 16th level of LiH and the 12th level of HF 
have nearly zero values, which indicate ``missing rungs'' in the vibrational levels.

To understand the cause of the MRP,
we start by considering a harmonic oscillator.
The vibrational wavefunctions of the harmonic potential $\Ket{\phi_i}$ are described by Hermite polynomials,
and their parities are different between adjacent levels, 
as described in Fig. \ref{fig:pes_tdm}\subref{fig:harm}. 
Assuming an odd function for the dipole moment,
the TDM between adjacent levels $\mu_{i,i+1}=\Bra{\phi_i}\mu\Ket{\phi_{i+1}}$
becomes an integral of an even function, 
which yields a nonzero value (i.e. an allowed transition).
The well-known infrared selection rule $\Delta \nu = \pm 1$ corresponds to this result \cite{willock2009molecular}.
On the other hand, if we assume an even function for the dipole moment,
the adjacent TDM $\mu_{i,i+1}$ becomes an integral of an odd function, 
which yields zero (i.e. a forbidden transition), 
as shown in the lower panel of Fig. \ref{fig:pes_tdm}\subref{fig:harm}.

Based on the above discussions, 
we consider two realistic molecular systems, LiH and HF,
where the shapes of the vibrational wavefunctions of molecules 
are similar to those of a harmonic potential 
except for a distortion due to the anharmonicity of the PESs, as shown in Figs. \ref{fig:pes_tdm}\subref{fig:LiH} and \subref{fig:HF}.
Therefore, the nature of the parity of the wavefunctions is approximately conserved,
whereas the parity of molecular dipole moments is not as simple as in the above discussion for the harmonic potential.
Near the equilibrium distance, 
the dipole moment can be regarded as being linear (i.e. odd function-like).
Thereby, the TDM $\mu_{i,i+1}$ increases monotonically 
as the vibrational level increases, as for the harmonic model. 
As the bond length increases, however, nonlinearity of the dipole moment appears.
The dipole function takes maxima (i.e. is even function-like) and then decreases asymptotically to zero.
This behavior results from the relaxation of the molecular polarization, 
which always occurs for charge neutral heteronuclear diatomic molecules.
As the edge of the distribution of the $i$th vibrational level of the wavefunction 
reaches the maxima of the dipole function,
the TDM $\mu_{i,i+1}$ begins to decrease 
due to the even-function nature of the dipole function.
Then the TDM has a near-zero value, 
as can be seen in Fig. \ref{fig:pes_tdm}\subref{fig:LiH} and \subref{fig:HF} 
for LiH and HF, respectively.
No missing rung will appear 
in the VLC simulations 
based on the assumption of the linear dipole function 
and is likely to give an unrealistic result.
Since the position of the missing rung 
sensitively depends on the shape of the PES and dipole function (see Supplemental Material), 
highly accurate quantum chemistry methods are required to identify its exact position.

\begin{figure*}[!t]
  \centering
  \subfloat[]{\includegraphics[keepaspectratio, scale=0.45]{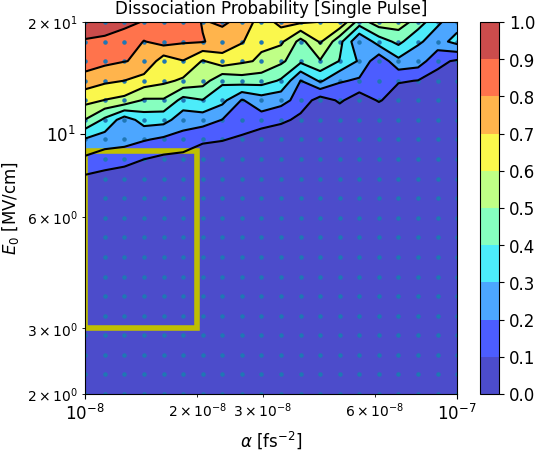}\label{fig:dissoc}}\hspace{1mm}
  \subfloat[]{\includegraphics[keepaspectratio, scale=0.45]{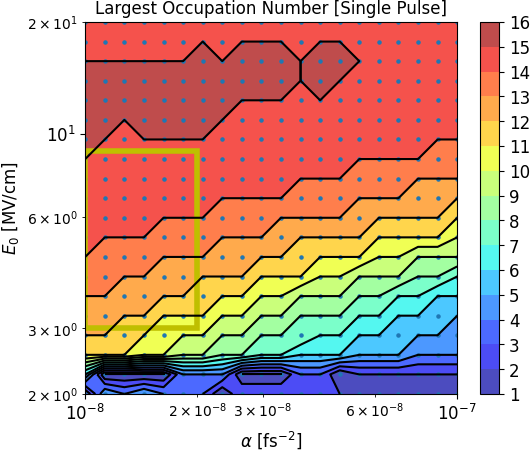}\label{fig:occu}}\hspace{1mm}
  \subfloat[]{\includegraphics[keepaspectratio, scale=0.43]{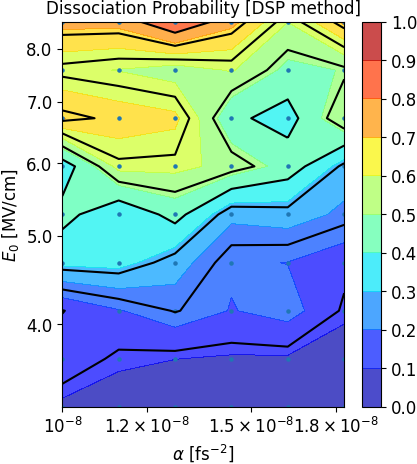}\label{fig:dsp}}
  \caption{Color maps of the results of the single pulse method and the DSP method (explained later in the paper).
          (a) Dissociation probabilities and (b) vibrational state with the largest occupation number for the single pulse method. 
          (c) Dissociation probabilities for the DSP method.
          The blue dots represent the grid points where the wavepacket simulations were conducted.
          The yellow boxed area in (a) and (b) represents the parameter range chosen for the DSP method.}
  \label{fig:map}
\end{figure*}

To verify the existence of the  missing rung 
by numerical experiments, 
we conducted wavepacket dynamics simulations for the LiH molecule. 
The time evolution of the wavepacket $\Psi (x,t)$ 
under a laser electric field $E(t)$ is expressed 
by the following time-dependent Schr\"{o}dinger equation
\begin{equation}
    i\hbar\frac{\partial}{\partial t}\Psi(x,t) = H(x,t)\Psi(x,t) ,
\end{equation}
where the Hamiltonian $H(x,t)$ is defined by 
\begin{align}
    H(x,t) &= H_0(x)-\mu(x) E(t) \\
    H_0(x) &= -\frac{\hbar^2}{2M_{\rm{red}}}\frac{\partial^2}{\partial x^2}+V(x) .
\end{align}
$M_{\rm{red}}$ in $H_0(x)$ is the reduced mass of LiH.
The potential energy $V(x)$ and the dipole moment $\mu(x)$ were calculated 
by the multireference averaged quadratic coupled-cluster (MR-AQCC) method \cite{szalay1993multi}.
Details of these computations are described in the Supplemental Material.
We employ a Gaussian pulse as the electric field $E(t)$ of the pulsed laser:
\begin{equation}
    E(t) = E_0\exp^{-\alpha (t-t_0)^2}\cos{\omega(t)(t-t_0)} ,
\end{equation}
where $E_0$ is the maximum electric field intensity, 
$\alpha$ is the Gaussian spreading parameter, 
$t_0$ is the center time of the pulse, 
and $\omega (t)$ is the time-dependent frequency.
The time variation of $\omega(t)$ is assumed to be a linear chirp, 
and parametrized by the two dimensionless parameters $\gamma_1$ and $\gamma_2$ as follows,
\begin{equation}
    \omega(t) = \omega_0\left\{ -(\gamma_1+\gamma_2) \cdot \frac{t-t_0}{4\sigma} + 1 + \frac{\gamma_1 - \gamma_2}{2} \right\} ,
\label{eq:omega}
\end{equation}
where $\sigma = {1}/{\sqrt{2\alpha}}$ is the standard deviation of the Gaussian pulse envelope.
Equation \ref{eq:omega} means that 
$\omega(t)$ is a linear function through two points, 
$\omega(t_0-2\sigma) = \omega_0(1+\gamma_1)$ and 
$\omega(t_0+2\sigma) = \omega_0(1-\gamma_2)$.
The reference frequency $\omega_0$ was set 
to the transition frequency between the ground state and the first excited state
($\omega_0 =  (E_1-E_0)/\hbar$).
Based on the above formulation, 
the VLC process for the LiH molecule was simulated 
by computing the time evolution of wavepackets 
initially set to the ground state 
using the second-order Suzuki--Trotter decomposition method.
Here, we discuss the parameter ranges of $E_0$ and $\alpha$ 
that determine the shape of the pulses.
As mentioned in the introduction, 
there are two types of excitation process mechanisms induced by chirped lasers: 
quantum ladder climbing and classical autoresonance. 
Since MRP is a quantum issue 
that apparently appears 
when the vibrational levels can be regarded as discrete, 
the parameter range should be chosen  
so that quantum ladder climbing occurs.
Therefore, using the characterization parameters 
proposed by Friedland et al. \cite{barth2011quantum}, 
we estimated the parameter range 
for quantum ladder climbing
(see Supplementary Material for details).
The parameter range was determined to be $E_0=2.0$--$20 [\rm{MV/cm}]$ and $\alpha=10^{-8}$--$10^{-7}[\rm{fs^{-2}}]$. 
For this parameter range of $E_0$ and $\alpha$, 
a grid of 20$\times$20 points on logarithmic scales was defined
and wavepacket dynamics simulations were carried out for each grid point. 
The chirp parameters $\gamma_1, \gamma_2$ were optimized by Bayesian optimization 
so as to maximize the degree of excitation and the dissociation probability.
The degree of excitation was evaluated
by the expectation value of the occupied states,
and the dissociation probability was evaluated 
by a time-integration of the probability density flux of the wavepacket 
that reaches the dissociation limit. 
The details of the wavepacket simulations and optimizations
are described in the Supplementary Material.

Figure \ref{fig:map}\subref{fig:dissoc} and \subref{fig:occu} shows 
the results of the wavepacket dynamics simulations 
with optimized chirp parameters.
The color map in Fig. \ref{fig:map}\subref{fig:dissoc} shows the dissociation probabilities, 
showing the trend that 
the dissociation probabilities get larger
with larger electric fields ($\propto E_0$) and longer pulse widths ($\propto 1/\sqrt{\alpha}$).
This result is a natural consequence of the fact 
that greater dissociation is promoted by a higher pulse energy.
The color map in Fig. \ref{fig:map}\subref{fig:occu} shows 
the vibrational state that has the largest occupation number (excluding the ground state) 
after each of the simulations.
It can be seen that there is a wide plateau consisting of the 15th and 16th states.
The missing rung of LiH is located around the 16th state, 
indicating that the wavepacket was trapped there due to the missing rung during the VLC.
This result shows that 
the MRP hinders photodissociation by VLC.
Since the MRP occurs for all molecular bonds with polarization, 
it is important to solve the MRP to achieve photodissociation by VLC.

Here, we propose the double-stepping pulse (DSP) method as a solution for the MRP.
In addition to a conventional pulse for the adjacent transitions ($\Delta \nu = \pm 1)$, 
the DSP method uses a secondary pulse 
to achieve transitions of $\Delta \nu = \pm 2$, i.e. double stepping, 
between levels around the missing rung.
Based on the previous discussion, 
when the dipole function has an even function property around the missing rung, 
transitions between levels with different parity ($\Delta \nu = \pm 1$) are forbidden, 
while transitions between levels with the same parity ($\Delta \nu = \pm 2$) are allowed,
and vice versa when the dipole function has an odd function property (see Supplementary Material).
Thereby, the DSP is complementary to a conventional pulse for successful VLC.

To show the effectiveness of our method, 
we conducted wavepacket dynamics simulations of the DSP method.
For the main pulse for the transitions of $\Delta \nu = \pm 1$ we chose pulse parameters for which almost no dissociation occurred using a single pulse
($E_0=3.0$--$9.0 [\rm{MV/cm}]$, $\alpha=1.0 \times 10^{-8}$--$2.0 \times 10^{-8} [\rm{fs^{-2}}]$, 
9 $\times$ 6 grid points on logarithmic scales,  
as shown in the yellow boxed area in Figs. \ref{fig:map}).
For the DSP, the pulse parameters $E_0=3.0 [\rm{MV/cm}]$, $\alpha = 8.0 \times 10^{-8} [\rm{fs^{-2}}]$ were used.
Here, we have to optimize the chirp parameters for two pulses, 
the main pulse and the DSP, 
and the delay time between the two pulses ($\Delta t_0 = t^{\rm{main}}_0-t^{\rm{DSP}}_0$).
Hence, we have to optimize five parameters, 
$\gamma^{\rm{main}}_1$, $\gamma^{\rm{main}}_2$, $\gamma^{\rm{DSP}}_1$, $\gamma^{\rm{DSP}}_2$, and $\Delta t_0$.
Since the computational cost of handling five variables in Bayesian optimization is high, 
the CMA-ES \cite{hansen1996adapting} evolutionary optimization method was used 
for the parameter optimization (see Supplementary Material for details).
The reference frequency of the DSP is set to the transition frequency between the 15th and 17th levels, 
($\omega_{0}^{DSP} = (E_{17} -E_{15})/\hbar$), 
where most of the wavepackets were trapped, as shown in Fig. \ref{fig:map}\subref{fig:occu}.
The relative phase between the main pulse and the DSP was set to zero.

\begin{figure}[bthp] 
  \centering
  \subfloat[]{\includegraphics[keepaspectratio, scale=0.35]{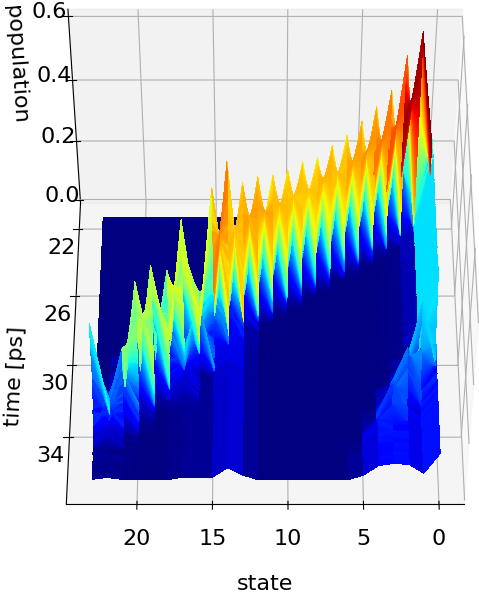}\label{fig:exc_with}}
  \subfloat[]{\includegraphics[keepaspectratio, scale=0.35]{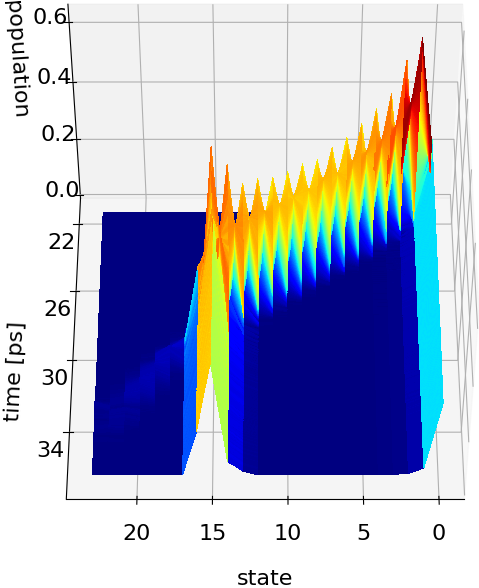}\label{fig:exc_without}}\\
  \subfloat[]{\includegraphics[keepaspectratio, scale=0.6]{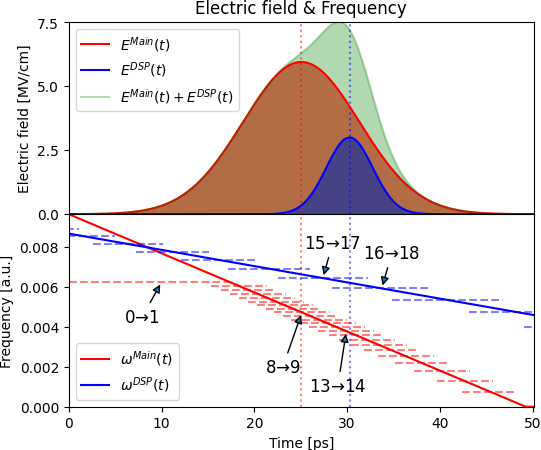}\label{fig:E_F}}
  \caption{(a) Snapshot of the excitation process with the most energy efficient pulses in the DSP method.
          The pulse parameters of the main pulse are $E_0=5.95[\rm{MV/cm}]$, $\alpha=1.27\times10^{-8}[\rm{fs^{-2}}]$.
          (b) Snapshot of the excitation process with only the main pulse, with the same parameters as in (a).
          (c) Upper panel: electric field intensities of the main pulse, the DSP and the overall field.
            Lower panel: corresponding time-dependent frequencies of the main pulse and the DSP in (a).
            The red and blue dashed lines represent the transition frequencies for $\Delta \nu = \pm 1$ (red) and $\Delta \nu = \pm 2$ (blue).}
  \label{fig:DSP_result}
\end{figure}

The dissociation probabilities obtained by the DSP method are shown in Fig. \ref{fig:map}\subref{fig:dsp}.
It can be seen that the DSP method enhances the dissociation significantly
compared to the single pulse method (Fig. \ref{fig:map}\subref{fig:dissoc}).
Although this result shows the superiority of the DSP method over the single-pulse method 
in terms of the dissociation probabilities, 
the energy efficiency of the DSP method remains a concern
because the method requires additional energy to generate the additional pulse.
Thus, we compared the energy efficiency 
of the two methods, specifically the amount of dissociated molecules per unit energy of the pulses.
(Details of the calculation method are described in the Supplementary Material.)
The highest energy efficiency for the DSP method was 0.763 [\rm{mol/J}]
while that for the single pulse method was 0.193 [\rm{mol/J}].
Thus, the DSP method was found to be about four times more efficient than the single pulse method, 
demonstrating that the DSP method is a promising method for solving the MRP.

Figure \ref{fig:DSP_result}\subref{fig:exc_with} shows 
a snapshot of the wavepacket dynamics simulations 
with the most energy efficient pulses in the DSP method. 
Figure \ref{fig:DSP_result}\subref{fig:exc_without} shows a snapshot 
with only the main pulse in Fig. \ref{fig:DSP_result}\subref{fig:exc_without}.
Without the DSP (Fig. \ref{fig:DSP_result}\subref{fig:exc_without}), 
it is clearly seen that 
the wavepacket is trapped in the missing rung level, 
while with the DSP (Fig. \ref{fig:DSP_result}\subref{fig:exc_with}), 
the wavepacket is successfully excited to the dissociation limit. 
Figure \ref{fig:DSP_result}\subref{fig:E_F} shows the electric fields and time-dependent frequencies 
used for the above simulation shown in Fig. \ref{fig:DSP_result}\subref{fig:exc_with}. 
It clearly shows that the DSP is focused on the 15th and 16th excitation frequencies 
which are around the missing rung. 

\begin{figure}[!b]
  \subfloat[]{\includegraphics[keepaspectratio, scale=0.4]{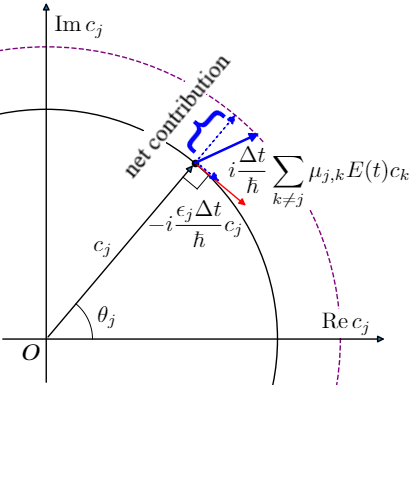}\label{fig:level}}
  \subfloat[]{\includegraphics[keepaspectratio, scale=0.55]{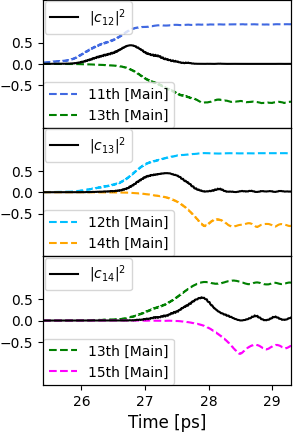}\label{fig:121314}}\\
  \caption{(a) Schematic illustration of Eq. \ref{eq:delta_prob_amp}.
          (b) Occupation numbers $|c_{j}|^2(t)$ and the contributions from other levels $\Delta C_{j}^{(k)}(t)$ 
          for the 12th, 13th, 14th levels.}
  \label{fig:contribution}
\end{figure}

However, the above discussions were limited to phenomenological cases. 
To understand the mechanism of the DSP method microscopically, 
we propose an analysis method that clarifies the dynamics of the vibrational levels
under electric fields.
Under an arbitrary electric field $E(t)$, 
the time evolution of the probability amplitude $c_j$ of the $j$th level, 
is expressed by the following equation, 
\begin{equation}
    i\hbar \frac{\partial c_j}{\partial t} = \epsilon_j c_j - \sum_{k \neq j} \mu_{j,k} E(t) c_k ,
    \label{eq:prob_amp}
\end{equation}
where $\epsilon_j$ is the eigenenergy of the $j$th level and $\mu_{j,k}$ is the TDM of the $j$th and $k$th levels.
Approximately, the variation of the probability amplitude $\Delta c_j$ during $\Delta t$ can be written as 
\begin{equation}
    \Delta c_j \simeq -i\frac{\epsilon_j\Delta t}{\hbar}c_j + i\frac{\Delta t}{\hbar}\sum_{k \neq j} \mu_{j,k} E(t) c_k .
    \label{eq:delta_prob_amp}
\end{equation}
Figure \ref{fig:contribution}\subref{fig:level} illustrates the meaning of Eq. \ref{eq:delta_prob_amp}.
The first term of the right hand side of Eq. \ref{eq:delta_prob_amp} 
does not change the norm of $c_j$
since it merely rotates $c_j$ in the complex plane.
On the other hand, the second term, 
which cause mixing with other levels via the electric field, 
affects the norm of $c_j$.
As shown in Fig. \ref{fig:contribution}\subref{fig:level}, 
the radial component of the second term changes the norm of $c_j$.
Therefore, the contribution to the norm of $c_j$ can be quantified 
through rotating $\Delta c_j$ by the argument of $c_j$, $\theta_j = \mathrm{arg} \, c_j$, and taking its real part.
The summation of the second term in Eq. \ref{eq:delta_prob_amp} represents the contributions to $c_j$ from each level.
We define $\Delta C_{j}^{(k)}(t)$ 
as the time integration of the contribution to the $j$th level from the $k$th level, 
indicating the net contribution of the $k$th level to the  $j$th level norm 
up to time $t$. 
It is written as
\begin{equation}
    \Delta C_{j}^{(k)}(t) = \int_{-\infty}^{t}  \mathrm{Re} \left( i\frac{\mu_{j,k} E(t')}{\hbar} c_k(t') \exp (-i\theta_j (t')) \right) dt' .
    \label{eq:contribution}
\end{equation}
A positive value of Eq. \ref{eq:contribution} means that 
the $k$th level acts to increase the norm of the $j$th level, 
and a negative value means that it acts to decrease it.
By calculating $\Delta C_{j}^{(k)}(t)$ for each level,
it is possible to analyze which vibrational level has the dominant contribution 
to the excitation process of a given level.
Furthermore, by calculating $\Delta C_{j}^{(k)}(t)$ 
for each electric field component $E^{\rm{Main}}(t)$ and $E^{\rm{DSP}}(t)$,
we can distinguish the contributions of each pulse.

Figure \ref{fig:contribution}\subref{fig:121314} shows 
the occupation number of the 12th, 13th, and 14th vibrational levels (levels lower than the missing rung)
and the contribution from the other levels 
calculated by Eq. \ref{eq:contribution} 
for the simulation of the DSP method shown in Fig. \ref{fig:DSP_result}\subref{fig:exc_with}.
It can be understood that the cascade excitation of a wavepacket, i.e. the VLC, is 
the result of quantum interference of the positive contribution from the lower level 
and the slightly delayed negative contribution from the upper level.
Note that 
these transitions are caused by contributions from adjacent levels, 
which agree with the process of quantum ladder climbing 
described by successive Landau--Zener transitions \cite{barth2011quantum}.
It is also worth noting that 
even though the electric field of the DSP is present 
during the transition of these levels, 
it does not contribute at all to the changes in occupation numbers.
This indicates that the excitation processes at levels lower than the MRP region 
are all induced by the main pulse.

\begin{figure}[bthp]
  \subfloat[]{\includegraphics[keepaspectratio, scale=0.5]{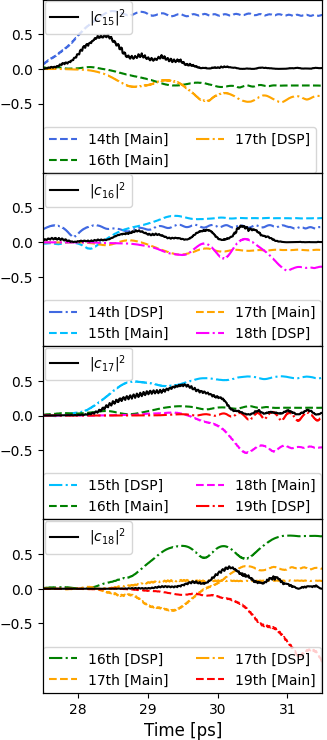}\label{fig:contrib_with}}
  \subfloat[]{\includegraphics[keepaspectratio, scale=0.5]{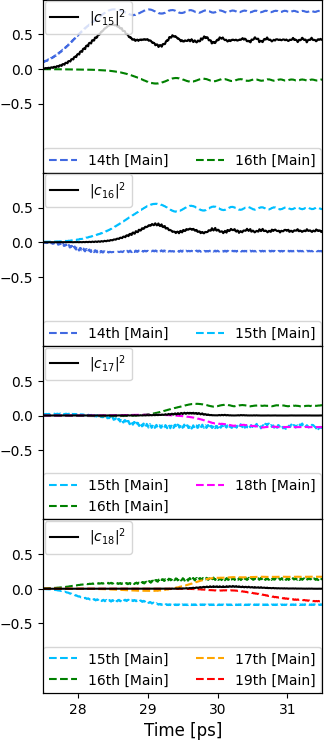}\label{fig:contrib_without}}\\
  \caption{Occupation numbers $|c_{j}|^2(t)$ and contributions from other levels $\Delta C_{j}^{(k)}(t)$ 
          for the 15th, 16th, 17th, and 18th levels 
          (a) with DSP and (b) without DSP.}
  \label{fig:level_dynamics}
\end{figure}
Next, 
we show the results for the vibrational levels of 
the 15th, 16th, 17th, and 18th (levels around the missing rung) in Fig. \ref{fig:level_dynamics}.
Figure \ref{fig:level_dynamics}\subref{fig:contrib_with} shows 
the result with DSP shown in Fig. \ref{fig:DSP_result}\subref{fig:exc_with}
and Fig. \ref{fig:level_dynamics}\subref{fig:contrib_without} shows 
the result without DSP shown in Fig. \ref{fig:DSP_result}\subref{fig:exc_without}.
Comparing the time variations of the occupation number for the 15th level,
it can be seen that 
without the DSP, 
the positive contribution of the 14th level is not sufficiently canceled 
by the negative contribution from the 16th level, and most of the wavepacket remains in the 15th level.
Whereas with the DSP, 
the additional negative contribution from the 17th level caused by the DSP 
significantly decreases the occupation number of the 15th level.
In the contribution to the 17th level in Fig. \ref{fig:level_dynamics} 
there are slight contributions from other levels without the DSP. 
On the other hand, there is an apparent positive contribution from the 15th level with the DSP, 
which excites wavepackets to higher levels.
Such an interlevel excitation can also be seen between the 16th and 18th levels for the DSP method.
These results confirm the interpretation of the MRP and the mechanism of the DSP method.

The proposed contributions reveal that 
the DSP method utilizes quantum interference effectively 
to solve the MRP.
Such quantum interference effects
are one of the most important features of coherent control \cite{ohmori2006real,katsuki2006visualizing,ohmori2003high}.
Another indication of quantum interference  
is the effects of the phases of the electric fields.
It is found that the dissociation probability changes significantly 
as the relative phase changes (see Supplemental Material). 
This implies that 
the phases of the electric field and that of the vibrational levels are coupled, 
and they play an important role in the dynamics of quantum interference.
Thus, quantum interference is deeply related to the DSP method.

In summary,
we simulated photodissociations by VLC
based on the PES and dipole moment
computed by a highly accurate quantum chemistry method. 
We found the MRP, which causes an interruption of VLC,
and revealed that the MRP is caused by the disappearance of the adjacent TDM at a specific vibrational level.
As a solution to this problem, 
we proposed the DSP method 
that uses additional pulses to induce a $\Delta \nu=\pm 2$ transition, 
and verified its effect on the MRP by wavepacket dynamics simulations. 
We note that 
the DSP method not only enhances the dissociation probability 
but is also more energy-efficient than the single pulse method.
We also clarified the detailed mechanisms of the VLC and DSP method. 
The excitation process of the VLC is caused by quantum interference 
between the positive contribution from the lower level and 
the negative contribution from the upper level 
and the DSP plays an essential role in the excitation process to higher levels.
The MRP is considered to be a ubiquitous problem in VLC 
because it originates in the parity of the vibrational wavefunction and the dipole function.
The MRP 
may be why there are few experimental reports of VLC-induced molecular bond breaking, 
and we believe that our findings may pave the way to VLC-induced photodissociation and, ultimately, 
to the versatile control of chemical reactions.

\bibliographystyle{unsrt}
\bibliography{main_ref.bib}

\begin{thebibliography}{10}

\bibitem{brumer1986control}
Paul Brumer and Moshe Shapiro.
\newblock Control of unimolecular reactions using coherent light.
\newblock {\em Chemical physics letters}, 126(6):541--546, 1986.

\bibitem{tannor1985control}
David~J Tannor and Stuart~A Rice.
\newblock Control of selectivity of chemical reaction via control of wave
  packet evolution.
\newblock {\em The Journal of chemical physics}, 83(10):5013--5018, 1985.

\bibitem{baumert1991femtosecond}
Thomas Baumert, M~Grosser, Rainer Thalweiser, and Gustav Gerber.
\newblock Femtosecond time-resolved molecular multiphoton ionization: The na 2
  system.
\newblock {\em Physical review letters}, 67(27):3753, 1991.

\bibitem{potter1992femtosecond}
ED~Potter, JL~Herek, S~Pedersen, Q~Liu, and AH~Zewail.
\newblock Femtosecond laser control of a chemical reaction.
\newblock {\em Nature}, 355(6355):66--68, 1992.

\bibitem{hikosaka2019coherent}
Y~Hikosaka, T~Kaneyasu, M~Fujimoto, H~Iwayama, and M~Katoh.
\newblock Coherent control in the extreme ultraviolet and attosecond regime by
  synchrotron radiation.
\newblock {\em Nature communications}, 10(1):1--5, 2019.

\bibitem{goto2011strong}
Haruka Goto, Hiroyuki Katsuki, Heide Ibrahim, Hisashi Chiba, and Kenji Ohmori.
\newblock Strong-laser-induced quantum interference.
\newblock {\em Nature Physics}, 7(5):383--385, 2011.

\bibitem{chelkowski1995adiabatic}
Szczepan Chelkowski and George~N Gibson.
\newblock Adiabatic climbing of vibrational ladders using raman transitions
  with a chirped pump laser.
\newblock {\em Physical Review A}, 52(5):R3417, 1995.

\bibitem{morichika2019molecular}
Ikki Morichika, Kei Murata, Atsunori Sakurai, Kazuyuki Ishii, and Satoshi
  Ashihara.
\newblock Molecular ground-state dissociation in the condensed phase employing
  plasmonic field enhancement of chirped mid-infrared pulses.
\newblock {\em Nature communications}, 10(1):1--8, 2019.

\bibitem{maas1997vibrational}
DJ~Maas, DI~Duncan, AFG Van~der Meer, WJ~Van~der Zande, and LD~Noordam.
\newblock Vibrational ladder climbing in no by ultrashort infrared laser
  pulses.
\newblock {\em Chemical Physics Letters}, 270(1-2):45--49, 1997.

\bibitem{marcus2006molecular}
Gilad Marcus, Arie Zigler, and Lazar Friedland.
\newblock Molecular vibrational ladder climbing using a sub-nanosecond chirped
  laser pulse.
\newblock {\em EPL (Europhysics Letters)}, 74(1):43, 2006.

\bibitem{jewariya2010ladder}
Mukesh Jewariya, Masaya Nagai, and Koichiro Tanaka.
\newblock Ladder climbing on the anharmonic intermolecular potential in an
  amino acid microcrystal via an intense monocycle terahertz pulse.
\newblock {\em Physical review letters}, 105(20):203003, 2010.

\bibitem{liu1995nonlinear}
Wing-Ki Liu, Binruo Wu, and Jian-Min Yuan.
\newblock Nonlinear dynamics of chirped pulse excitation and dissociation of
  diatomic molecules.
\newblock {\em Physical review letters}, 75(7):1292, 1995.

\bibitem{liu1999classical}
W-K Liu, J-M Yuan, and SH~Lin.
\newblock Classical dynamics of multiphoton excitation and dissociation of
  diatomic molecules by infrared laser pulses.
\newblock {\em Physical Review A}, 60(2):1363, 1999.

\bibitem{duan2000classical}
Yiwu Duan, Wing-Ki Liu, and Jian-Min Yuan.
\newblock Classical dynamics of ionization, dissociation, and harmonic
  generation of a hydrogen molecular ion in intense laser fields: A collinear
  model.
\newblock {\em Physical Review A}, 61(5):053403, 2000.

\bibitem{marcus2004quantum}
G~Marcus, L~Friedland, and A~Zigler.
\newblock From quantum ladder climbing to classical autoresonance.
\newblock {\em Physical Review A}, 69(1):013407, 2004.

\bibitem{barth2011quantum}
I~Barth, L~Friedland, O~Gat, and AG~Shagalov.
\newblock Quantum versus classical phase-locking transition in a
  frequency-chirped nonlinear oscillator.
\newblock {\em Physical Review A}, 84(1):013837, 2011.

\bibitem{barth2014quantum}
Ido Barth and Lazar Friedland.
\newblock Quantum phenomena in a chirped parametric anharmonic oscillator.
\newblock {\em Physical review letters}, 113(4):040403, 2014.

\bibitem{armon2017quantum}
Tsafrir Armon and Lazar Friedland.
\newblock Quantum versus classical dynamics in the optical centrifuge.
\newblock {\em Physical Review A}, 96(3):033411, 2017.

\bibitem{armon2019quantum}
Tsafrir Armon and Lazar Friedland.
\newblock Quantum versus classical effects in the chirped-drive discrete
  nonlinear schr{\"o}dinger equation.
\newblock {\em Physical Review A}, 100(2):022106, 2019.

\bibitem{armon2020quantum}
Tsafrir Armon and Lazar Friedland.
\newblock Quantum versus classical chirps in a rydberg atom.
\newblock {\em Physical Review A}, 102(5):052817, 2020.

\bibitem{barth2015ladder}
Ido Barth, Ilya~Y Dodin, and Nathaniel~J Fisch.
\newblock Ladder climbing and autoresonant acceleration of plasma waves.
\newblock {\em Physical review letters}, 115(7):075001, 2015.

\bibitem{shalibo2012quantum}
Yoni Shalibo, Ya’ara Rofe, Ido Barth, Lazar Friedland, Radoslaw Bialczack,
  John~M Martinis, and Nadav Katz.
\newblock Quantum and classical chirps in an anharmonic oscillator.
\newblock {\em Physical review letters}, 108(3):037701, 2012.

\bibitem{lin1998quantum}
JT~Lin, TL~Lai, DS~Chuu, and Tsin-Fu Jiang.
\newblock Quantum dynamics of a diatomic molecule under chirped laser pulses.
\newblock {\em Journal of Physics B: Atomic, Molecular and Optical Physics},
  31(4):L117, 1998.

\bibitem{mishima1998theoretical}
Kenji Mishima and Koichi Yamashita.
\newblock A theoretical study on laser control of a molecular nonadiabatic
  process by ultrashort chirped laser pulses.
\newblock {\em The Journal of chemical physics}, 109(5):1801--1809, 1998.

\bibitem{witte2003controlling}
Thomas Witte, Thomas Hornung, Lars Windhorn, Detlev Proch, Regina
  de~Vivie-Riedle, Marcus Motzkus, and Karl-Ludwig Kompa.
\newblock Controlling molecular ground-state dissociation by optimizing
  vibrational ladder climbing.
\newblock {\em The Journal of chemical physics}, 118(5):2021--2024, 2003.

\bibitem{willock2009molecular}
David Willock.
\newblock {\em Molecular symmetry}.
\newblock John Wiley \& Sons, 2009.

\bibitem{szalay1993multi}
P{\'e}ter~G Szalay and Rodney~J Bartlett.
\newblock Multi-reference averaged quadratic coupled-cluster method: a
  size-extensive modification of multi-reference ci.
\newblock {\em Chemical physics letters}, 214(5):481--488, 1993.

\bibitem{hansen1996adapting}
Nikolaus Hansen and Andreas Ostermeier.
\newblock Adapting arbitrary normal mutation distributions in evolution
  strategies: The covariance matrix adaptation.
\newblock In {\em Proceedings of IEEE international conference on evolutionary
  computation}, pages 312--317. IEEE, 1996.

\bibitem{ohmori2006real}
Kenji Ohmori, Hiroyuki Katsuki, Hisashi Chiba, Masahiro Honda, Yusuke Hagihara,
  Katsutoshi Fujiwara, Yukinori Sato, and Kiyoshi Ueda.
\newblock Real-time observation of phase-controlled molecular wave-packet
  interference.
\newblock {\em Physical review letters}, 96(9):093002, 2006.

\bibitem{katsuki2006visualizing}
Hiroyuki Katsuki, Hisashi Chiba, Bertrand Girard, Christoph Meier, and Kenji
  Ohmori.
\newblock Visualizing picometric quantum ripples of ultrafast wave-packet
  interference.
\newblock {\em Science}, 311(5767):1589--1592, 2006.

\bibitem{ohmori2003high}
Kenji Ohmori, Yukinori Sato, Evgueni~E Nikitin, and Stuart~A Rice.
\newblock High-precision molecular wave-packet interferometry with hgar dimers.
\newblock {\em Physical review letters}, 91(24):243003, 2003.

\end{thebibliography}


\begin{thebibliography}{10}

\bibitem{szalay1993multi}
P{\'e}ter~G Szalay and Rodney~J Bartlett.
\newblock Multi-reference averaged quadratic coupled-cluster method: a
  size-extensive modification of multi-reference ci.
\newblock {\em Chemical physics letters}, 214(5):481--488, 1993.

\bibitem{roos1980complete}
Bj{\"o}rn~O Roos, Peter~R Taylor, and Per~EM Sigbahn.
\newblock A complete active space scf method (casscf) using a density matrix
  formulated super-ci approach.
\newblock {\em Chemical Physics}, 48(2):157--173, 1980.

\bibitem{dunning1989gaussian}
Thom~H Dunning~Jr.
\newblock Gaussian basis sets for use in correlated molecular calculations. i.
  the atoms boron through neon and hydrogen.
\newblock {\em The Journal of chemical physics}, 90(2):1007--1023, 1989.

\bibitem{kendall1992electron}
Rick~A Kendall, Thom~H Dunning~Jr, and Robert~J Harrison.
\newblock Electron affinities of the first-row atoms revisited. systematic
  basis sets and wave functions.
\newblock {\em The Journal of chemical physics}, 96(9):6796--6806, 1992.

\bibitem{prascher2011gaussian}
Brian~P Prascher, David~E Woon, Kirk~A Peterson, Thom~H Dunning, and Angela~K
  Wilson.
\newblock Gaussian basis sets for use in correlated molecular calculations.
  vii. valence, core-valence, and scalar relativistic basis sets for li, be,
  na, and mg.
\newblock {\em Theoretical Chemistry Accounts}, 128(1):69--82, 2011.

\bibitem{schmidt1993general}
Michael~W Schmidt, Kim~K Baldridge, Jerry~A Boatz, Steven~T Elbert, Mark~S
  Gordon, Jan~H Jensen, Shiro Koseki, Nikita Matsunaga, Kiet~A Nguyen, Shujun
  Su, et~al.
\newblock General atomic and molecular electronic structure system.
\newblock {\em Journal of computational chemistry}, 14(11):1347--1363, 1993.

\bibitem{gordon2005advances}
Mark~S Gordon and Michael~W Schmidt.
\newblock Advances in electronic structure theory: Gamess a decade later.
\newblock In {\em Theory and applications of computational chemistry}, pages
  1167--1189. Elsevier, 2005.

\bibitem{huber1979constants}
Klaus-Peter Huber.
\newblock Constants of diatomic molecules.
\newblock {\em Molecular spectra and molecular structure}, 1979.

\bibitem{van2007accurate}
W~Van~Dijk and FM~Toyama.
\newblock Accurate numerical solutions of the time-dependent schr{\"o}dinger
  equation.
\newblock {\em Physical Review E}, 75(3):036707, 2007.

\bibitem{barthel2020optimized}
Thomas Barthel and Yikang Zhang.
\newblock Optimized lie--trotter--suzuki decompositions for two and three
  non-commuting terms.
\newblock {\em Annals of Physics}, 418:168165, 2020.

\bibitem{barth2011quantum}
I~Barth, L~Friedland, O~Gat, and AG~Shagalov.
\newblock Quantum versus classical phase-locking transition in a
  frequency-chirped nonlinear oscillator.
\newblock {\em Physical Review A}, 84(1):013837, 2011.

\bibitem{kemlin2016transient}
Vincent Kemlin, Adeline Bonvalet, Louis Daniault, and Manuel Joffre.
\newblock Transient two-dimensional infrared spectroscopy in a vibrational
  ladder.
\newblock {\em The journal of physical chemistry letters}, 7(17):3377--3382,
  2016.

\bibitem{snoek2012practical}
Jasper Snoek, Hugo Larochelle, and Ryan~P Adams.
\newblock Practical bayesian optimization of machine learning algorithms.
\newblock {\em Advances in neural information processing systems}, 25, 2012.

\bibitem{gpy2014}
{GPy}.
\newblock {GPy}: A gaussian process framework in python.
\newblock \url{http://github.com/SheffieldML/GPy}, since 2012.

\bibitem{lawrence2003fast}
Neil Lawrence, Matthias Seeger, and Ralf Herbrich.
\newblock Fast sparse gaussian process methods: The informative vector machine.
\newblock In {\em Proceedings of the 16th annual conference on neural
  information processing systems}, number CONF, pages 609--616, 2003.

\bibitem{hansen1996adapting}
Nikolaus Hansen and Andreas Ostermeier.
\newblock Adapting arbitrary normal mutation distributions in evolution
  strategies: The covariance matrix adaptation.
\newblock In {\em Proceedings of IEEE international conference on evolutionary
  computation}, pages 312--317. IEEE, 1996.

\bibitem{hansen2016cma}
Nikolaus Hansen.
\newblock The cma evolution strategy: A tutorial.
\newblock {\em arXiv preprint arXiv:1604.00772}, 2016.

\bibitem{loshchilov2013bi}
Ilya Loshchilov, Marc Schoenauer, and Mich{\`e}le Sebag.
\newblock Bi-population cma-es agorithms with surrogate models and line
  searches.
\newblock In {\em Proceedings of the 15th annual conference companion on
  Genetic and evolutionary computation}, pages 1177--1184, 2013.

\bibitem{foote2017ionization}
David~B Foote, Y~Lin, Liang-Wen Pi, JM~Ngoko Djiokap, Anthony~F Starace, and
  WT~Hill~III.
\newblock Ionization enhancement and suppression by phase-locked ultrafast
  pulse pairs.
\newblock {\em Physical Review A}, 96(2):023425, 2017.

\end{thebibliography}

\end{document}


\title{Supplemental Material for ``Missing Rung Problem in Vibrational Ladder Climbing"}

\author{Takahiro Horiba}
\author{Soichi Shirai}
\author{Hirotoshi Hirai}
\email{hirotoshih@mosk.tytlabs.co.jp}

\affiliation{Toyota Central Research and Development Labs., Inc., 41-1, Yokomichi, Nagakute, Aichi 480-1192, Japan}

\date{\today}

\maketitle

\section{Quantum chemistry computations}
\begin{figure}[th!]
    \centering
    \subfloat[][]{\includegraphics[keepaspectratio,scale=0.5]{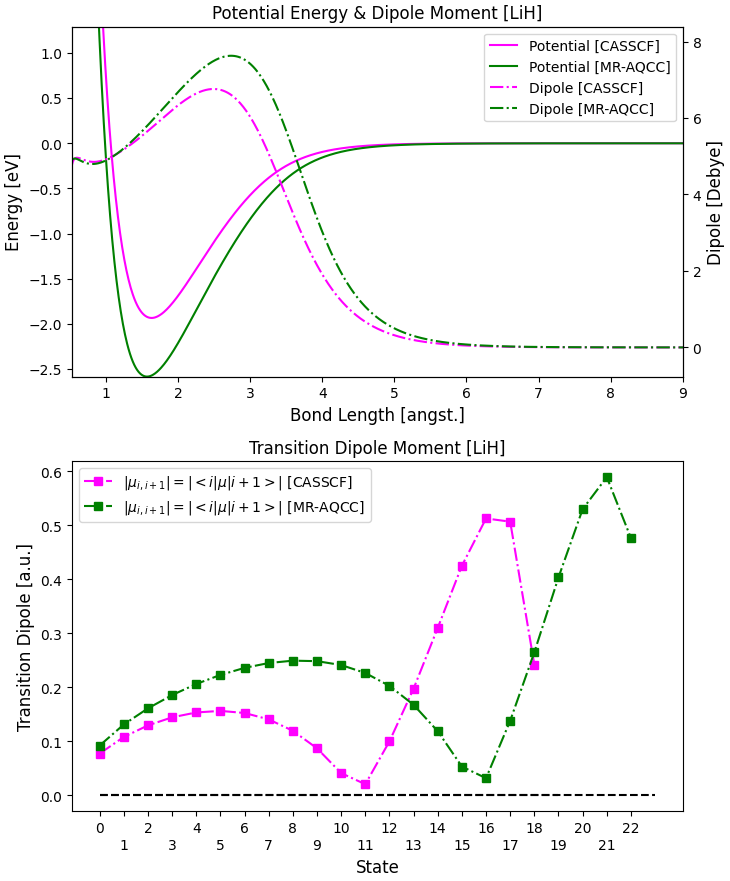}\label{fig:LiH_pes}}
    \subfloat[][]{\includegraphics[keepaspectratio,scale=0.5]{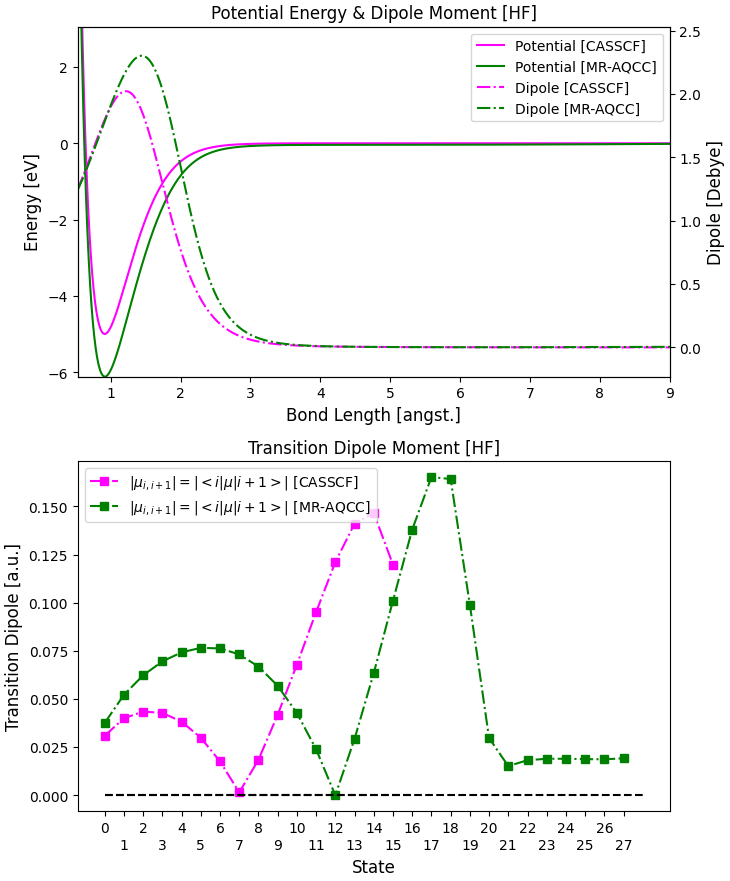}\label{fig:HF_pes}}
    \caption{Comparison between CASSCF and MR-AQCC. Upper panels: PESs and dipole moments for (a) LiH molecule and (b) HF molecule. Lower panels: absolute values of the corresponding TDMs.}
    \label{fig1_sm}
\end{figure}

The potential energy curves and dipole moments of the LiH and
HF molecules were calculated using the
multi-reference averaged quadratic coupled-cluster (MR-AQCC) method \cite{szalay1993multi}.
First, complete active space self-consistent field (CASSCF) calculations \cite{roos1980complete}
were carried out and the obtained CASSCF wavefunctions were adopted as
reference functions for the MR-AQCC calculations.
For LiH, the molecular orbitals derived from the Li 2s and H 1s atomic orbitals
were selected as active orbitals. The CAS was constructed by distributing
two electrons over these two orbitals. In the MR-AQCC calculations,
the electrons in Li 1s were additionally correlated.
For HF, the molecule was placed on the z-axis. Accordingly, the H 1s and F 2pz
orbitals were relevant for the formation of the H-F bond. The molecular orbitals
derived from these atomic orbitals were selected as active orbitals
in the CASSCF calculations; two electrons involved in these orbitals were
treated as active electrons. The electrons in the orbitals originating from
F 1s, F 2s, F 2px, and F 2py were also correlated in the MR-AQCC calculations.
The basis set used was Dunning's aug-cc-pVQZ \cite{dunning1989gaussian,kendall1992electron,prascher2011gaussian}.
All the calculations were carried out using the GAMESS program \cite{schmidt1993general,gordon2005advances}.
The calculated spectroscopic parameters and dipole moments are
summarized in Table I. Here, the spectroscopic parameters were estimated
by fitting the potential energy curve with a quadratic function. 
The calculated parameters were in good agreement with the corresponding
experimental values \cite{huber1979constants}. The results suggest that
the potential energy curves and dipole moments are reliable over the entire range.
The CASSCF and MR-AQCC potential energy curves and transition dipole moments are compared in Fig. \ref{fig1_sm}.
The CASSCF results were quite different from the MR-AQCC results because of the lack of dynamical correlations.
As a result, the missing rung in which the transition dipole moment is close to zero
appeared at different states. The results suggest that calculations using a high-level method
are necessary to estimate the exact position of the missing rung.
\begin{table}[h!]
  \caption{Calculated and experimental parameters for LiH and HF. $r_e$ are the equilibrium distances, $D_0$ are the dissociation energies, $\omega_e$ are the fundamental vibrational frequencies and $\mu$ are the dipole moments. }
  \label{table1}
  \centering
  \begin{tabular}{c|c|c|c|c}
    \hline
    System  & $r_e$ [angst.]  & $D_0$ [eV]  & $\omega_e$ [cm$^{-1}$] & $\mu$ [Debye]\\
    \hline
    LiH (calc.) & 1.5710 & 2.496& 1422.22 & 5.792 \\
    LiH (exp.) & 1.5957 & 2.429 & 1405.65 & 5.882 \\
    HF (calc.) & 0.9168 & 5.848 & 4103.48 & 1.791 \\
    HF (exp.) & 0.9168& 5.869 & 4138.32 & 1.826 \\
    \hline
  \end{tabular}
\end{table}

\section{Double-stepping transition}
\begin{figure}
    \subfloat[][]{\includegraphics[keepaspectratio, scale=0.8]{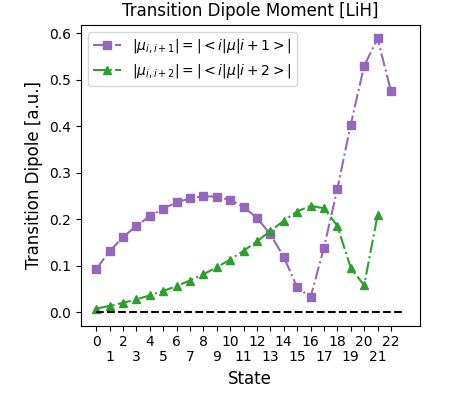}\label{fig:LiH}}
    \subfloat[][]{\includegraphics[keepaspectratio, scale=0.8]{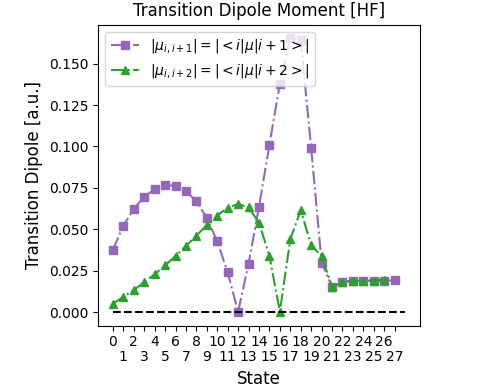}\label{fig:HF}}
    \caption{Absolute values of TDMs for $\Delta \nu = \pm 1$ and TDMs for $\Delta \nu = \pm 2$ for the (a) LiH molecule and (b) HF molecule.
            The TDMs are calculated based on the PESs and dipole moments calculated by MR-AQCC.}
    \label{fig:ssp_dsp}
\end{figure}

Here, we show the transition dipole moments (TDMs) of $\Delta \nu = \pm 2$.
Figures \ref{fig:ssp_dsp}\subref{fig:LiH} and \subref{fig:HF} show 
the TDMs of $\Delta \nu = \pm 1$ transitions and $\Delta \nu = \pm 2$ transitions of LiH and HF molecules.
It is clearly seen that the increase and decrease of these values are opposite, 
which can be explained by the parities of wavefunctions and the dipole moment.

\section{Wavepacket dynamics simulations}
Non-relativistically, the time evolution of quantum systems can be described by the time-dependent Schr\"{o}dinger equation,
\begin{equation}
    i\hbar\frac{\partial}{\partial t}\psi(x, t) = H(t) \psi (x, t).
\end{equation}
The formal solution of the Schr\"{o}dinger equation with the time-independent potential $V(x)$ can be written as
\begin{equation}
\psi(x, t) = e^{-i \frac{H(x)}{\hbar} t}\psi (x, 0),
\end{equation}
where  $U(t) = e^{-i \frac{H(x)}{\hbar} t}$ is the time evolution operator.
However, we have to consider the time-dependent Hamiltonian $H(x, t)$ that describes the system under the laser electric field $E(t)$,
\begin{equation}
H(x, t) = -\frac{\hbar^2}{2M}\frac{\partial^2}{\partial x^2} + V(x, t),
\end{equation}
where $V(x, t) = V(x) - \mu(x) E(t)$.
To treat the time-varying potential $V(x, t)$, we set $t = N dt$ and express the time evolution operator as the product of $N$ operators in time increments of $dt$,
\begin{equation}
U(t) = \prod_{j = 0}^{N} U_j(dt),
\end{equation}
where
\begin{equation}
U_j(dt) = e^{-i \frac{H(x, jdt)}{\hbar}dt}
\end{equation}

Several methods of applying an exponential operator to a wave function have been developed \cite{van2007accurate}, including methods based on a Taylor expansion 
and the Chebyshev-polynomial method. However, 
the Suzuki--Trotter decomposition method \cite{barthel2020optimized} is often used because of the good balance between computational cost and speed. 
We also use the Suzuki--Trotter decomposition method for the time-evolution computation.
The second-order Suzuki--Trotter decomposition can be written as
\begin{equation}
U_j(dt) \sim e^{-i \frac{T}{\hbar} \frac{dt}{2}}e^{-i \frac{V(x, jdt+dt/2)}{\hbar} dt}e^{-i \frac{T}{\hbar} \frac{dt}{2}} +O(dt^3).
\end{equation}
An error arises because $T$ and $V$ are non-commutative. 
Here, $T$ is diagonal in wavenumber space and $V$ is diagonal in real space, so in the second-order Suzuki--Trotter method,
when applying $e^{-i \frac{T}{\hbar} \frac{dt}{2}}$, the wave function is expressed in wavenumber space using a fast Fourier transformation (FFT)
and $e^{-i \frac{V(x, jdt+dt/2)}{\hbar} dt}$ is applied to the wave function, which returns to real space notation by an inverse FFT. 
When $e^{-i \frac{T}{\hbar} \frac{dt}{2}}$ is applied again, the wave function is expressed in wave number space using the FFT again. 
Using the above procedure, the time-evolution computation can proceed without expanding the exponential function operator.
By repeating these calculations $N$ times, the wave function at the desired time $t$ can be obtained. 

For the simulations of LiH dissociation by chirped laser pulses, $dt = 1.0$ [a.u.] is used.
The simulation time $t$ depends on the shape of the chirped laser pulses, $t = 8\sigma$,
where  $\sigma$ is the standard deviation of the Gaussian envelope for the pulse laser.
We take an evenly spaced isotropic grid of $2^{10}$ elements for $x$ in the range $x=1.5$--$15.0$ [bohr].

\subsection{Dissociation probability}
Here, we describe how to calculate the dissociation probability.
The wavefunctions that reach the right edge of the grid represent the state of bond-dissociation.
Therefore, the probability densities $|\psi(x_{right-edge})|^2$ represent the dissociation probability.
It is convenient to use the probability density flux,
\begin{equation}
J(x, t) = \frac{\hbar}{2iM}(\psi^*(x, t)\nabla \psi(x, t) -(\nabla \psi(x, t))^*  \psi(x, t)) = Re(\psi(x, t)^*\frac{\hbar}{iM}\nabla \psi(x, t)).
\end{equation}
$\frac{\hbar}{iM}\nabla \psi(x, t)$ can be transformed as the following,
\begin{equation}
\frac{\hbar}{iM}\nabla \psi(x, t) = \hat{p}\psi(x,t)/M = IFFT(k\psi(p, t))/M,
\end{equation}
where $IFFT$ means an inverse FFT.
Thereby we can quickly compute $\frac{\hbar}{iM}\nabla \psi(x, t)$ because we have the wavefunction in wave number space $\psi(p, t)$ and can therefore use the Suzuki--Trotter decomposition method for the time-evolution computation.
The time integral of the probability density flux at the right edge of the grid after a sufficiently long simulation time gives the dissociation probability in this case:
\begin{equation}
P_{dissociation} = \int_o^t J(x_{right-edge}, t')dt'.
\end{equation}

\subsection{Complex absorbing potential}
An absorption potential with a negative pure imaginary value was set to prevent the wavefunction from reaching the grid boundary.
Here, we used the complex absorption potential of a 4th order function,
\begin{equation}
V_{absorbing} (x) = -i \xi x^4
\end{equation}
where $\xi = 1.0$ is used in this study.
The norm of the wave function reaching the region where this function has a significant value decays.
This prevents reflections at the edge and the consequent unwanted interference.
We placed the potential just after the flux calculation point.

\section{Range of Laser Parameters}
Since the MRP is a quantum problem that appears 
when the vibrational levels can be regarded as being discrete, 
we have to choose laser parameters ($E_0$: maximum electric field intensity, 
$\alpha$: Gaussian spreading parameter)
that reproduce the excitation process 
where quantum phenomena are dominant.
Here, we provide a rough estimation of the laser parameter range 
where quantum ladder climbing occurs, 
using the parameters proposed by Friedland et al. 
to distinguish between quantum ladder climbing and classical autoresonance \cite{barth2011quantum}.

The excitation process is characterized by three time scales, 
$T_R=\sqrt{2m\hbar \omega_0}/\epsilon$, 
$T_S=1/\sqrt{\Gamma}$,
and $T_{\rm{NL}}=2\omega_0 \beta / \Gamma$
where $m$ is the reduced mass, 
$\omega_0$ is the eigenfrequency, 
$\beta$ is the anharmonicity of the system,
$\epsilon$ is the magnitude of external field, 
and $\Gamma$ is the chirp rate.

To obtain the values of $\omega_0$ and $\beta$ of LiH, 
the excitation frequency is fitted with the following equation:
\begin{equation}
    \omega_{n,n+1} = \omega_0[1-2\beta(n+1)].
\end{equation}
The fitting gives $\omega_0=0.0068 [\rm{a.u.}]$, $\beta = 0.0176$.

$\epsilon$ corresponds to the product of the electric field 
and the linear dipole moment.
Although the dipole moment of LiH is nonlinear, 
in the range 2--5.5 [Bohr], it is linearly approximated 
to be $\mu(x) = 0.354x + 1.086$.
The effective external field magnitude $\epsilon^{\rm{eff}}$ 
is estimated as $\epsilon^{\rm{eff}}=0.354E_0$.
The chirp rate $\Gamma$ is taken as the value 
when $\gamma_1=\gamma_2=0.5$, namely $\Gamma=\omega_{0 \to 1}/4\sigma$.

Based on these values, 
$T_R$, $T_S$, and $T_{\rm{NL}}$ are calculated. 
The parameters that distinguish the quantum case from the classical case $P_1$, $P_2$ are expressed as follows, 
\begin{align}
    P_1 &= \frac{T_S}{T_R} ,\\
    P_2 &= \frac{T_{\rm{NL}}}{T_S} .
\end{align}

The conditions under which phase-locked ladder climbing occurs are as follows, 
\begin{align}
    P_2 &> P_1 + 1 ,\\
    P_1 &> 0.79 .
    \label{eq:condition}
\end{align}

\begin{figure}
    \centering
    \includegraphics[keepaspectratio, scale=0.8]{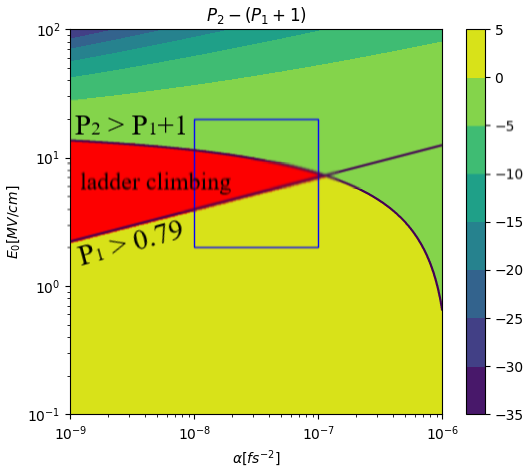}
    \caption{Characterization parameter $P_2 - (P_1+1)$ calculated in the ranges $E_0=10^{-1}$--$10^2$, $\alpha=10^{-9}$--$10^{-6}$ .
    The parameter range corresponding the quantum ladder climbing is colored red.
    The blue boxed area represents the parameter ranges we chose.}
    \label{fig:friedland}
\end{figure}

Figure \ref{fig:friedland} shows the calculated value of $P_2 - (P_1+1)$ 
for the parameter ranges $E_0=10^{-1}$--$10^2$, $\alpha=10^{-9}$--$10^{-6}$.
The parameter range that satisfy the conditions Eq. \ref{eq:condition} 
is colored in red.
From Fig. \ref{fig:friedland}, 
ladder climbing is considered to occur 
in the parameter range $E_0=2.0$--$20$ [\rm{MV/cm}], $\alpha < 10^{-7}$ [\rm{fs}$^{-2}]$.
It is not possible to make $\alpha$ arbitrarily small 
due to the vibrational relaxation time.
We refer to the vibrational lifetime of the CO stretching measured by Kemlin et al. \cite{kemlin2016transient}  
and assume a time scale for vibrational relaxation of a few tens of ps, 
which corresponds to the parameter range $\alpha= 10^{-8}$--$10^{-7}[\rm{fs}^{-2}]$.
Based on the above estimation, 
we set the parameter ranges as 
$E_0=2.0$--$20$ [\rm{MV/cm}], $\alpha= 10^{-8}$--$10^{-7}[\rm{fs}^{-2}]$ 
(blue boxed area in the Fig. \ref{fig:friedland}).
Note that 
a more detailed discussion is required for an actual system 
to obtain a more accurate characterization.

\section{Optimization of the chirped laser pulse parameters}
Here, we describe the details of the optimization methods of chirped laser parameters.
We adopted the following linear chirped frequency 
as the time-dependent frequency $\omega(t)$ of the laser pulse,
\begin{equation}
    \omega(t) = \omega_0\left\{ -(\gamma_1+\gamma_2) \cdot \frac{t-t_0}{4\sigma} + 1 + \frac{\gamma_1 - \gamma_2}{2} \right\}.
    \label{eq:td-frequency}
\end{equation}
Equation \ref{eq:td-frequency} is parameterized by the two dimensionless parameters $\gamma_1$ and $\gamma_2$.
These chirp parameters significantly affect the excitation process of VLC.
To achieve a more effective excitation, 
we optimized these parameters 
using a combination of machine learning and a wavepacket dynamics simulation.
We used machine learning to update the parameters 
based on the evaluated values of the parameters obtained by the wavepacket dynamics simulation.

For a single pulse, 
the two parameters $\gamma_1$ and $\gamma_2$ are optimized 
in a range restricted to $[0,1]$ so as to represent the downchirp.
This parameter space is divided into a 100 $\times$ 100 grid (i.e. 0.01 steps), 
and the optimal parameter is searched from these grid points by the optimization method.

The chirp parameters are optimized for each of 400 different pulses, 
thus a fast and efficient optimization method is required. 
In addition, 
since the effects of the parameters on the excitation process cannot be evaluated analytically, 
this system is regarded as a black-box.
As a fast black-box optimization method, 
we employed Bayesian optimization.
Bayesian optimization enables efficient exploration and optimization in parameter space 
leveraging predictive models such as Gaussian process regression, 
which is a powerful method for black-box optimization \cite{snoek2012practical}.
We used Gpy \cite{gpy2014}, an open source library in Python, for the implementation of Bayesian optimization.
In the optimization, 
we adopted Gaussian process regression with the RBF (radial basis function) kernel 
as the predictive model for the parameter space, 
and employed the UCB (upper confidential bound)
as the acquisition function.
The iteration number for updating parameters was set to 50.

The objective function was set to be 
the sum of the excited and dissociated wavepackets.
Although we aim to find parameters that give efficient photodissociation, 
for some pulse parameters whose energy is not sufficient to cause dissociation, 
an objective function with only an amount of dissociation 
will result in insufficient optimization.
Therefore, by including an amount of an excited wavepacket 
in the objective function in addition to the dissociated wavepacket, 
it is possible to obtain optimized parameters that realize efficient excitation.
The amount of dissociated wavepacket is evaluated 
by integrating the absorbed wavepacket described in the previous section.
The degree of excitation is evaluated by the expected value of the occupied states $\Ket{\phi_i}$, as follows, 
\begin{equation}
    P_{\rm{excited}} = \sum_{i=0}^{n_{\rm{dissoc}}}\left( \frac{i}{n_{\rm{dissoc}}} \right) |\Braket{\phi_i|\psi(x,t)}|^2 ,
\end{equation}
where $n_{\rm{dissoc}}$ is the highest level in the bound state, 
$n_{\rm{dissoc}}=23$ in the case of LiH.
The expected value of the level is normalized by $n_{\rm{dissoc}}$ 
in order to balance the amount of dissociation, whose value range is $[0,1]$.
Based on the above settings, 
optimization of the chirp parameters was performed for a single pulse.

In the case of the DSP method, 
there are five parameters to be optimized, 
the chirp parameters of the main pulse and the DSP,
$\gamma^{\rm{Main}}_1$, $\gamma^{\rm{Main}}_2$, $\gamma^{\rm{DSP}}_1$, and $\gamma^{\rm{DSP}}_2$,
and the delay time between the main pulse and the DSP ($\Delta t_0 = t^{\rm{DSP}}_0-t^{\rm{Main}}_0$).

Since the computational cost of Bayesian optimization increases 
in proportion to the cube of the number of variables \cite{lawrence2003fast}, 
it is difficult to handle five variables in Bayesian optimization.
Thus, CMA-ES \cite{hansen1996adapting} was used for optimization.
CMA-ES 
adaptively updates and optimizes the covariance matrix of the multivariate normal distribution 
that generates candidate solutions based on the evaluation values of the candidates.
It is known that CMA-ES is a robust method 
that can be optimized even in noisy high-dimensional parameter spaces \cite{hansen2016cma,loshchilov2013bi}.

In the optimization using CMA-ES, 
the number of generations was set to 150 
and the number of individuals to 16 for each generation.
The objective function was set to be only the dissociated wavepacket 
because a certain amount of dissociation can be expected by adding the DSP.
The initial values $\gamma^{\rm{Main}}_1$, $\gamma^{\rm{Main}}_2$ are inherited 
from the optimal parameters for the single pulse case.
The value range of the parameters is not restricted, 
allowing the parameters to be negative.
Negative parameters give the possibility of upchirping or 
the DSP reaching ahead of the main pulse.

\section{Estimation of Energy Efficiency}
To determine the energy efficiency of photodissociation by VLC, 
the energy of the laser pulse and
the amount of dissociated molecules
have to be estimated.
Here, we show a rough estimation of these values.
First, we explain how to estimate the energy of a pulse.
The electric field of a Gaussian pulsed laser $E(t)$ is determined 
by the maximum electric field intensity $E_0$ and 
the Gaussian spreading parameter $\alpha$ 
as follows,
\begin{equation}
    E(t) = E_0\exp^{-\alpha (t-t_0)^2}\cos{\omega(t)(t-t_0)}.
    \label{eq:electric_field}
\end{equation}
The irradiance $I[\rm{W/cm^2}]$ of a pulsed laser 
is expressed using the maximum electric field intensity $E_0[\rm{V/cm}]$, as follows \cite{foote2017ionization},
\begin{equation}
    I = \frac{c\epsilon_0 n}{2}E^2_0 ,
    \label{eq:irradiance}
\end{equation}
where $c$, $\epsilon_0$, and $n$ are 
the speed of light the dielectric constant, and refractive index of the vacuum, respectively.
The energy of a pulse $e[\rm{J}]$ is expressed 
by the irradiance $I[\rm{W/cm^2}]$, pulse width $T[\rm{s}]$, and cross section of the pulse $S[\rm{cm^2}]$ as follows, 
\begin{equation}
    e = STI \simeq  0.002 \cdot \frac{E_0^2 S}{\sqrt{\alpha}} ,
    \label{eq:energy}
\end{equation}
where the pulse width $T[s]$ is defined 
as the full width at half maximum of the Gaussian pulse, $T=2\sqrt{2\log{2}}\sigma \simeq 2.23/\sqrt{2\alpha}$.
In the case of a single pulse, 
the total energy required to cause VLC is calculated by Eq.\ref{eq:energy}.
In the case of the DSP method,
two pulses, the main pulse and the DSP, 
are required for VLC. 
We assume that the optical system 
for the implementation of the DSP method 
is like that used in pump-probe spectroscopy with a single light source.
Assuming that there is no energy loss in such an optical system, 
the total energy required for VLC become the sum of the main pulse and the DSP energy.
Since this is a very rough estimate, 
it is considered that more energy is required for the DSP method in practice. 

Next, we explain the calculation method 
for the amount of dissociated molecules.
The amount of molecules $N_{\rm{all}}[\rm{mol}]$, 
present in the region of the pulse cross section $S[\rm{cm^2}]$, 
can be written as $N_{\rm{all}} = \rho S$, 
where $\rho[\rm{mol/cm^2}]$ is the areal density of the molecules.
Assuming that the laser pulse causes dissociation 
with a dissociation probability $d$, 
the amount of dissociated molecules $N_{\rm{dissoc}}[\rm{mol}]$ is given by 
\begin{equation}
    N_{\rm{dissoc}}=N_{\rm{all}}d=\rho S d .
    \label{eq:dissoc}
\end{equation}
For simplicity, we assume that $\rho = 1 [\rm{mol/cm^2}]$.

From Eqs. \ref{eq:energy} and \ref{eq:dissoc}, 
the energy efficiency $P[\rm{mol/J}]$ of photodissociation by VLC 
can be estimated as follows, 
\begin{equation}
    P = \frac{N_{\rm{dissoc}}}{e} \simeq 478 \cdot \frac{d \sqrt{\alpha}}{E_0^2} .
    \label{eq:efficiency}
\end{equation}
Based on the above formulation, 
we estimated the energy efficiency 
for the results of the single-pulse case and DSP method, respectively.
\begin{figure}[tbhp]
    \subfloat[]{\includegraphics[keepaspectratio,scale=0.65]{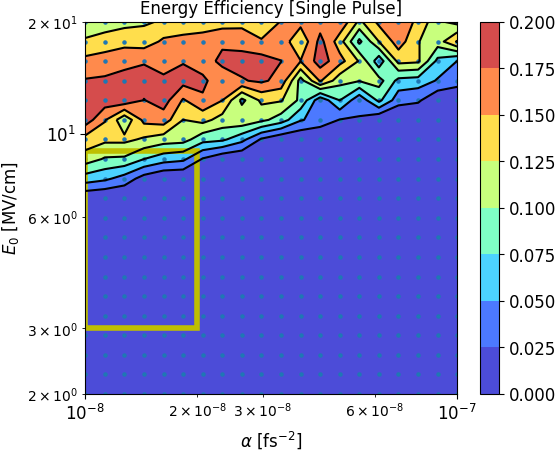}}\hspace{1mm}
    \subfloat[]{\includegraphics[keepaspectratio,scale=0.65]{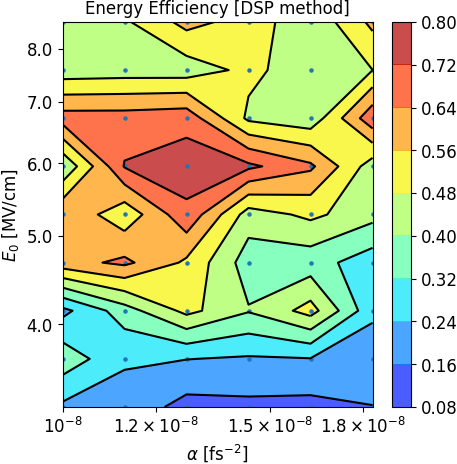}}
    \caption{Calculated energy efficiencies for (a) a single pulse and (b) the DSP method}
    \label{fig:ene_effi}
\end{figure}

Figure \ref{fig:ene_effi} shows the calculated energy efficiency of each case.
It can be seen that 
the DSP method shows a higher energy efficiency 
than the single pulse case.
The maximum energy efficiency for a single pulse is 0.193 [\rm{mol/J}],
while that for the DSP method is 0.763 [\rm{mol/J}],
which is about four times higher.
This result shows the superiority of the DSP method 
with respect to energy efficiency.

\section{Effect of Relative Phase}
In the optimization of the parameters of the DSP method, 
the relative phase 
between the electric field of the main pulse and that of the DSP 
was fixed at zero.
This is because the relative phase changes significantly 
with changes in the parameters, such as the chirp parameters, 
making it difficult to optimize it simultaneously with those parameters.
However, the effect of the relative phase is not negligible, 
since it affects the time evolution of the probability amplitudes of the vibrational levels.
In this section, 
we discuss the effect of the relative phase between the main pulse and DSP 
on the photodissociation by the DSP method.

For the 54 pulses considered in the DSP method with optimized parameters at a relative phase of zero, 
we performed a wavepacket dynamics simulation by varying the relative phases from 0 to $2\pi$, 

\begin{figure}[bt]
    \centering
    \includegraphics[width=0.4\linewidth]{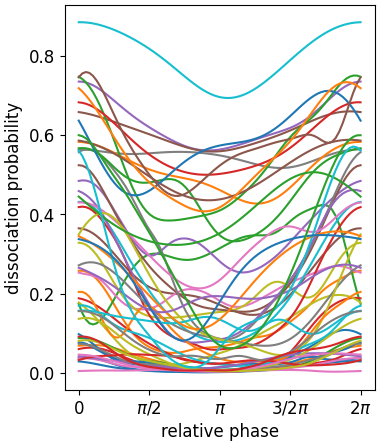}
    \caption{Relationship between the dissociation probability and relative phases.}
    \label{fig:relative_phase}
\end{figure}

Figure \ref{fig:relative_phase} shows the relationship 
between the dissociation probability and relative phase.
Figure \ref{fig:relative_phase} 
shows that the relative phase has a significant effect on the dissociation probability.
The change in relative phase affects 
the phase of the probability amplitude of the vibrational levels 
and changes the interference between them, 
which leads to a change in the dissociation probability, shown in Fig.\ref{fig:relative_phase}.
Therefore, this result can be attributed to 
the quantum interference effect between the vibration levels.

In addition, it is notable that 
the relationship between the dissociation probability and the relative phase 
is approximately cosine functional.
Since the optimization was performed with the relative phase fixed at zero, 
the optimized parameters were considered to have been chosen 
to maximize the dissociation probability at a phase of zero.
The fact that the dissociation probability becomes minimum at a relative phase of $\pi$, 
that is the antiphase to the case of zero, 
is a characteristic result in interference phenomena.
Since the excitation process results from the complex interference 
between many vibrational levels,
it is not obvious that such a cosine functional relationship appears 
between the dissociation probability and relative phase.
A complete interpretation of these results will require a more detailed discussion.

\clearpage

\bibliographystyle{unsrt}
\bibliography{supplement_ref}